\documentclass[journal]{IEEEtran}
\usepackage{amsmath,graphicx,color}
\usepackage{cite}
\usepackage{hyperref}
\usepackage{amsthm}
\usepackage{algorithm}
\usepackage{algorithmic}
\usepackage{amsfonts}
\usepackage{url}
\usepackage{booktabs}
\usepackage{multirow} 
\usepackage{stfloats}
\usepackage{flushend}
\newcommand{\orcid}[1]{\href{https://orcid.org/#1}{ {\includegraphics[scale=0.5]{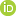}}}}
\title{Low-Rank Cyclostationarity Predictive Routing Is Almost as Good as Real-Time Data-based Routing}
\author{Oriel~Singer$^{1,2}$\textsuperscript{\orcid{0009-0003-1546-8925}},
        Ilai~Bistritz$^{2}$\textsuperscript{\orcid{0000-0002-4120-8292}},~\IEEEmembership{Member,~IEEE}
        Giseung~Park$^{3}$\textsuperscript{\orcid{0000-0002-9737-4142}},
        Woohyeon~Byeon$^{4}$\textsuperscript{\orcid{0009-0004-2993-9297}},
        Youngchul~Sung$^{4}$\textsuperscript{\orcid{0000-0003-4536-6690}},~\IEEEmembership{Senior~Member,~IEEE}
        and~Amir~Leshem$^{1}$\textsuperscript{\orcid{0000-0002-2265-7463}},~\IEEEmembership{Fellow,~IEEE}
\thanks{$^1$ Bar-Ilan University, Israel.
$^2$ Tel Aviv University, Israel. $^3$ University of Toronto, Canada. $^4$ KAIST, Republic of Korea.\\
Corresponding author: Amir Leshem, amir.leshem@biu.ac.il \\
O. Singer and A. Leshem were partially supported by the Israel Ministry of Innovation, 
Science \& Technology, grant no. 1001556423 and the ISF grant 2197/22.
This work was also supported by the National Research Foundation of Korea (NRF) grant funded by the Korea government (MSIT) (2022K1A3A1A31093462). \\
We have used ChatGPT 5.2 and Claude Opus 4.5 to review and improve the text and coding the simulation. All simulations were verified independently.}}

\begin{document}
\maketitle
\begin{abstract}
Dynamic shortest-path routing, using real-time traffic data, enables path selection responsive to evolving conditions. Nevertheless, transportation planning tasks such as adaptive congestion pricing, fleet routing, and long-term operational decisions rely on offline traffic estimators. To address this problem, we develop a spatiotemporal predictor based on a low-rank decomposition of the traffic matrix and the temporal subspace coefficients. Using a recent large-scale measurement campaign over the Seoul road network, we show that our proposed predictor incurs an average excess travel time of less than 1.5 minutes. Moreover, our predictor's tail of the excess travel time distribution matches that of a near-real-time predictor. Results based on one year of traffic data are also demonstrated in simulations.
\end{abstract}

\section{Introduction}
\IEEEPARstart{M}{ any} transportation decisions must be made before traffic conditions are realized. For example, mechanisms such as congestion pricing, scheduled tolling schemes or fleet dispatch policies are computed in advance and are intended to influence future route choices rather than react to them. Such interventions cannot rely on instantaneous measurements of the full network state. Instead, they require predicting the network-wide travel-time structure in advance.

Since drivers primarily seek to minimize travel time, the key quantity to predict is the shortest-path structure. Reliable prediction of network-wide travel times would allow planners to anticipate driver routing behavior and make informed decisions without requiring real-time data.

In this paper, we use a new unique empirical graph data collected from Seoul, of the average travel time in $\sim$5000 arterial roads, at 10-minutes sampling rate on average, collected from May 2023 to June 2024. This dataset is unique in that it covers the main traffic arteries of a large metropolitan area over a one-year period, which is significantly longer than what is typically available in public repositories for large-scale traffic networks of this size.

Based on this large dataset, we propose a novel spatiotemporal model for forecasting link weights. We show that, with the dynamic shortest path algorithm, 
our simple offline prediction model relying on past data yields only a minor difference in travel time for almost all cases compared to real-time data predicted shortest routes, practically obviating the need of real-time data for routing, and opens many new opportunities to efficient traffic system management.      

\section{Related Work}
Kim et al. \cite{kim2005optimal} compared historical and real-time routing on several freeways, but measured benefits in fleet operating costs rather than individual travel-time regret at the city scale.  

Other works combine prediction and routing directly. Lee et al. \cite{lee2024transfer} aimed to improve routing with RL-based guidance, but tested only on small synthetic grids. Lagos et al. \cite{lagos2025online} treated routing as an online learning problem, comparing against a fixed best path rather than real-time baselines.  
Chen et al. \cite{chen2013finding} examined reliable paths with large Hong Kong data, focusing on on-time arrival probabilities rather than regret. Zhang and Li \cite{zhang2023finding} used real-time measurements at departure to drive an online predictor and modeled its error distributions for stochastic routing, but did not evaluate how much of the routing gap can be closed by an offline prediction without any real-time input. Falek et al. \cite{falek2022re} showed that given real-time data over real networks, using static (no reroute) Dijkstra performs almost the same as dynamic Dijkstra with real-time data using, but did not compare it to any prediction model.

Low-rank and decomposition approaches \cite{asif2016matrix,mitrovic2015low,lin20243d} established that traffic data can be represented compactly with spatial modes that can recover missing data or allow large-scale compressed sensing. These works retrain or re-estimate the decomposition per dataset or time window, whereas we fix the spatial basis from an initial slice and reuse it for restoring and predicting later periods. Wang et al. \cite{wang2021graphtte} estimated end-to-end path travel time on road network graphs using spatiotemporal GCN-GRU models, but assumed a fixed trajectory is given rather than selecting the optimal path.

Compact spatiotemporal representations such as contrastive learning, graph-signal, and autoencoders were studied \cite{li2024sts,jin2024spatio,zhang2024arfa}.   
The survey in \cite{ermagun2018spatiotemporal} highlights the importance of spatial dependence in forecasting and the lack of full-network models. 

Deep learning approaches \cite{yu2017spatio,xu2025fddsgcn,wu2025sfadnet,guo2019deep} can improve forecasting accuracy. However, our objective is not minimizing prediction error per se, but rather predicting weights with sufficient fidelity to support reliable shortest-path routing, both on average and in the tail. Our simple predictor achieves this, matching the routing performance of near-real-time benchmarks, by exploiting low-rank spatial structure and cyclic temporal correlations rather than complex nonlinear architectures.

\textbf{Our contributions are threefold}:
\begin{enumerate}
    \item We introduce a large-scale, year-long dataset of travel times 
    over $\sim$5,000 arterial road segments in Seoul.
    \item We propose a novel simple predictor combining cyclostationary 
    temporal dynamics with low-rank spatiotemporal modes.
    \item We show that our predictor nearly matches dynamic real-time routing, suggesting that offline prediction can substitute for real-time data in transportation planning.
\end{enumerate}

\section{problem formulation}

We consider a weighted graph of roads with time-varying weights, where the weight of an edge is the travel time on that road. Let $G(t) = (V, E, W(t))$ be a discrete time-dependent graph with dynamic weights updated at fixed intervals.  

Let $(e_1, e_2, \dots , e_n) \subseteq E $ be the ordered edge sequence included in a given path from $u$ to $v$ at start time $t_\text{start}$. The total time to complete the path would be:
\begin{equation}
  T = \sum_{i=1}^n w_{e_i,t_i}, 
\end{equation}
where $w_{e_i,t}$ is the required time  to travel edge $e_i$, entering  edge $e_i$ at time $t$, and 
\begin{equation}
t_i = t_1 + \sum_{j=1}^{i-1} w_{e_j,t_j}~~~\mbox{with}~t_1 = t_{\text{start}}.
\end{equation}

We assess the predicted weights by applying a dynamic shortest path algorithm both on predicted weights and real-time weights and comparing their travel times.   
Specifically, let $T_{rt}(I)$ and $T_{pred}(I)$ be the dynamic shortest path travel time using real-time data and predicted data, respectively, for input $I = [G(V,E), t_{start}, u, v]$. Then the predictor's regret is defined as the extra travel-time taken over the specified input: 
\begin{equation}
    R(I) = T_{pred}(I) - T_{rt}(I).
\end{equation}

The mean regret can obscure less frequent but consequential failures. A predictor with low average regret may still exhibit a heavy tail, leading to occasional but substantial excess travel times that are unacceptable in practice. To capture this risk, we emphasize the tail of the regret distribution via its Complementary CDF (CCDF) and upper $\alpha$-quantile, focusing on low-probability, high-impact events:
\begin{equation}
    q^{up}(\alpha) \;=\; \; \inf \Bigl\{ y : \Pr(\text{R} > y ) \leq \alpha \Bigr\}.
\end{equation}
This tail-oriented evaluation highlights the robustness of routing decisions and directly reflects the extent to which a predictor may expose drivers to severe delays. 

\section{Spatiotemporal Predictor}
In this section we present our predictor for network-wide link weights.  Our predictor is based on a spatiotemporal model that combines a low-rank representation of the spatial structure with a cyclostationary modeling of temporal dynamics.  
\textbf{Data and spatiotemporal basis.}  
Let $\mathbf{W} \in \mathbb{R}^{m \times n}$ be the average travel-time matrix with $m$ road segments, i.e., edges of the graph, and $n$ time intervals, so the $(i,t)$-th element of $\mathbf{W}$ is the average travel time of segment $i$ at time $t$. Each column of $\mathbf{W}$ corresponds to the weight vector of the road network graph $G$ at a specific time interval. To capture essential spatial correlation and obtain a spatial basis, we 
apply the singular-value decomposition on the average travel-time matrix $\mathbf{W}$:
\begin{equation}
        \mathbf{W} =  \mathbf{U}  \mathbf{\Sigma} \mathbf{\Xi}^\top.
\end{equation}
We retain the first $k ~(\ll m \ll n)$ left singular-vectors of $\mathbf{W}$, i.e., the first $k$-columns of $\mathbf{U}$   
as our spatial basis $\bar{\mathbf{U}} \in \mathbb{R}^{m \times k}$. This captures stable spatial correlations between segments, which are assumed to persist over time. Thus, the spatial basis $\bar{\mathbf{U}}$ is fixed once estimated and reused as new data arrives.  

We model the temporal dynamics as {\em cyclostationary}: that is, patterns repeat from cycle to cycle (e.g., daily or weekly), while variation occurs within each cycle. Then, our predictor is defined by a cycle period and an intra-cycle resolution. Specifically, let the length of a cycle be  \texttt{CyclePeriod} and the estimation time resolution within each cycle be  \texttt{Resolution}. So, the number of intervals per cycle is computed by $L = {\text{CyclePeriod}}/{\text{Resolution}}$. Then, for each interval $l \in \{1,\dots,L\}$, our model estimates a coefficient vector $\boldsymbol \alpha_{l} \in \mathbb{R}^k$ that modulates the spatial basis $\bar{\mathbf{U}}$ to yield an estimate of weight vector, capturing the estimated travel times for all road segments.  

\textbf{Estimation of the modulation vector.}  
To capture potential minor temporal shifts along cycles, we update the modulation vector when a new cycle of data
\[\mathbf{w}_{l,p}=[w_{l,p}^{(1)},\cdots, w_{l,p}^{(m)}]^\top,~l=1,\cdots,L\] arrives, where $p$ is the index for the new cycle. The new observations from each interval $l$  are projected onto $\bar{\mathbf{U}}$, yielding the coefficient vector $\hat{\boldsymbol{\alpha}}_{l,p}$ via least-squares:
\begin{equation}
\hat{\boldsymbol\alpha}_{l,p}
= \mathop{\arg\min}_{\mathbf{\alpha}}\,\big\|\mathbf{w}_{l,p}-\bar{\mathbf{U}}\boldsymbol{\alpha}\big\|_2^2,~~~\forall l.
\end{equation}
Then,  new vectors are aggregated into a running mean across cycles. That is, for prediction at cycle $c$, the modulation vector $\boldsymbol{\alpha}_{l,c-1}$ is the average of all estimates from cycles $1,\dots,c-1$, which is accomplished by   
 the running mean:
\begin{equation}    
\boldsymbol\alpha_{{l,p}}
= \frac{p-1}{p}\,\boldsymbol\alpha_{{l,p-1}}
  + \frac{1}{p}\,\hat{\boldsymbol\alpha}_{{l,p}}, ~~~p=1,\cdots, c-1
\end{equation}
with $\boldsymbol{\alpha}_{l,-1}=\mathbf{0}$. Note that this moving average filter averages all $\hat{\boldsymbol{\alpha}}_{l,p},~p=1,\cdots,c-1$ with equal weights.

\textbf{Prediction and efficiency.}  Finally, our predictor for all road segments at interval $l$ of cycle $c$ is given by 
\begin{equation}
        \hat{\boldsymbol w}_{l,c} = \bar{\mathbf{U}} \boldsymbol\alpha_{{l,c-1}},
\end{equation}
Hence, our model reconstructs the expected travel-time profile at a given interval by combining the global mean with spatial factors and cycle-specific temporal weights.  

Note that our prediction model is memory-efficient: instead of storing all past modulation  vectors, it maintains only the running averages $\boldsymbol{\alpha}_{l,c-1}$ for each interval and basis dimension, storing only $L \cdot k$ values in addition to the spatial basis $\bar{\mathbf{U}}$.

\textbf{Spatial Space Analysis}
The MDL model order estimator Wax and Kailath \cite{wax1985detection} attains its minimum at 73 modes. However, the MDL score becomes nearly flat around 25 modes (see Supplementary Materials). Since the performance of the $k=25$ is similar but implementation has much lower complexity (factor 9) we prefer to use the order $k=25$.   

\textbf{Time Space Analysis}
We used Welch periodogram \cite{stoica2005spectral} to estimate the power spectral density (PSD) of the right singular vectors from the truncated SVD of the temporal domain. The results reveal a clear dominance of daily and weekly frequencies along the first dozen temporal singular vectors analyzed. The spectral analysis of the singular vectors appears in the Supplementary Materials.

\textbf{Dynamic Routing Shortest Path Algorithm}
The dynamic shortest path approach repeatedly applies a static shortest-path algorithm at each decision point (vertex) using the updated weights at the time of arrival to that point. This optimal causal routing strategy was used for all the predictors, including the real-time routing benchmark.

\begin{algorithm}[!t]
\caption{Greedy Re-routing Shortest Path}
\label{alg:greedy_rerouting}
\begin{algorithmic}[1]
\REQUIRE $I = [G(V,E), t, u, v]$ where $u, v \in V$
\ENSURE List of edges forming predicted path from $u$ to $v$

\STATE Initialize \texttt{edges} $\gets [\,]$, \texttt{nodes} $\gets [u]$
\STATE Set $t_{curr} \gets t$, $u_{curr} \gets u$
\WHILE{$u_{curr} \neq v$}
    \STATE Predict all edges travel time $\hat{\mathbf{w}}_{t_{curr}}$
    \STATE Run Dijkstra's algorithm on $G$ from $u_{curr}$ to $v$ using predicted weights $\hat{\mathbf{w}}_{t_{curr}}$, to obtain a path
    \STATE Append $e^*$, the first edge in the path, to \texttt{edges}
    \STATE Append $u^*$, the endpoint of $e^*$, to \texttt{nodes}
    \STATE Update $t_{curr} \gets t_{curr} + \textit{traveltime}(e^*)$ and $u_{curr} \gets u^*$
\ENDWHILE
\RETURN \texttt{edges}
\end{algorithmic}
\end{algorithm}

\section{Numerical Results}
\subsection{Preprocessing the data}
The raw dataset reported the average travel times (sec) every $\sim$10 minutes for 5,048 sensor-defined segments, each identified by GPS start and end coordinates ($\sim$0.1 m accuracy).

\textbf{Data alignment and segment-wise imputation:} 
 We standardized all recordings to a fixed 10-minute grid. Values not within 5 minutes of a target slot were linearly interpolated using the closest points within $\pm$10 minutes; short gaps of up to two intervals were also interpolated. Outliers- less than one-fifth of a segment’s mean travel time were removed.

\textbf{Graph construction:}
Segment endpoints were projected to Seoul’s CRS and snapped (STRtree with 100 m threshold) to junctions in the OpenStreetMap (OSM) graph \cite{openstreetmap,boeing2017osmnx}. To avoid redundant OSM nodes (e.g., multiple points per intersection), nearby nodes ($<$100 m) were clustered with DBSCAN, and their centroids used as junctions. We retained only the largest strongly connected component, yielding 4,729 valid segments and 1,731 junctions.

\textbf{Graph level imputation:} We temporally interpolated segments that had up to 2 consecutive missing samples. About $5.5\%$ of data were missing after interpolation and discarding long blackouts ($>3$ h, and over $30\%$). For this missing data, we performed a spatio-temporal interpolation using neighboring segments to recover the missing data.
\subsection{Training and test sets}
We constructed two predictors based on our model: one incorporating a daily cycle and the other a weekly cycle. Both predictors use a low-rank representation consisting of 25 spatial basis vectors, trained on the first month of data starting in May 2023. 
We tested larger numbers of basis vectors, but observed only negligible improvements, indicating that the additional dimensions primarily capture noise. 
As baselines, we constructed dynamic one-day-lag and dynamic one-week-lag predictors, which reuse the edge weights from the same time on the previous day or the previous week, respectively, applied in a re-routing manner. We also evaluated a dynamic 10-minute-lag predictor, which approximates the dynamic real-time benchmark, and a static shortest path solution, which represents optimal performance without re-routing capability.

To build a representative test set, we applied Louvain community detection \cite{blondel2008fast} to identify closely connected areas (22 communities found), classifying them as “inner” (city center) or “outer” (periphery). Origin–destination (OD) pairs were then formed by sampling all inner–outer and outer–inner community combinations, with nodes randomly chosen within each community. This yielded \textbf{2863 OD pairs}.

For each OD, we sampled roughly \textbf{13 timestamps per day} (rounded hours between 6 AM–7 PM). The test period spans the 13 months (June 2023–June 2024) after the one-month training set (May 2023), excluding several short gaps and one longer gap of about a month in December 2023. After discarding routes with (real) travel time below 30 minutes\footnote{For shorter routes the regret was negligible}, our final test set included \textbf{4.6 million OD–time inputs}.

\subsection{Analysis and comparison}
Our cyclostationary predictors substantially improved over lagged baselines: mean regret of 1.23 and 1.38 minutes compared to 2.32 and 2.52 minutes for one-week-lag and one-day-lag, respectively, as seen in Table \ref{table_regret}.  More importantly, our models nearly matched the performance of benchmarks using near-real-time data (i.e., 10-minute-lag: 1.15 minutes) and actual realized conditions (static: 0.97 minutes).

While mean regret is an overall performance measure, the tail behavior of routing errors is paramount for reliability and user experience. We examined tail-regret across multiple spatio-temporal partitions (workday vs. weekend, hour of day, day of week, traffic direction), and found that the predictor hierarchy remained remarkably consistent across all domains. Our cyclostationary predictors consistently outperformed one-day-lag and one-week-lag baselines, with the weekly-cycle variant performing slightly better than the daily-cycle model. All predictors exhibited degraded performance during afternoon hours (3-7pm) due to increased real-time traffic variance.

Figure 1 shows that even the 10-minute-lag and static shortest path predictors, which leverage near-current or real-time traffic conditions, experience significant tail-regret, \textbf{indicating unpredictable real-time traffic fluctuations}. In  $\sim$25\% of trips, our offline predictor outperforms the real-time benchmark. This occurs because greedy re-routing may commit to a suboptimal detour based on transients, after which the remaining path to the destination is inevitably longer.

\begin{figure*}[!t]
    \centering
    \includegraphics[width=0.95\textwidth]{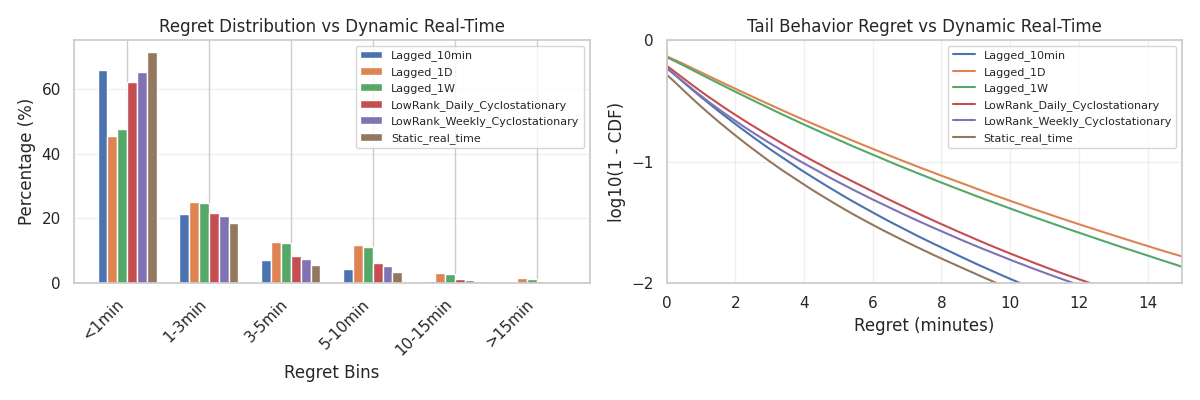}
    \caption{Distribution and tail behavior of routing regret. Left: Cyclostationary predictors (red/purple) exhibit regret distributions closely matching 10-minute-lag (blue) and static (brown) benchmarks. Right: log CCDF of tail regret on log scale.}
    \label{fig:regret}
\end{figure*}

Importantly, our lowrank-cyclostationary models exhibit tail behavior nearly aligned with the 10-minute lag predictors, demonstrating that our approach captures the predictable component of traffic dynamics. Table \ref{table_regret} compares our model's tail regret statistics vs the near-optimal predictors. 
\begin{table}[!t]
\centering
\caption{Regret Statistics (Minutes) vs. Dynamic Real-Time}
\label{table_regret}
\begin{tabular}{lccc}
\toprule
\textbf{Predictor} 
  & \textbf{Mean} 
  & \shortstack{\textbf{10\% upper} \\ \textbf{quantile}} 
  & \shortstack{\textbf{1\% upper} \\ \textbf{quantile}} \\
\midrule
1-day lag & 2.52 & 6.94 & 17.71 \\
1-week lag & 2.32 & 6.49 & 16.50 \\
Low-rank daily cycle& 1.38 & 4.31 & 12.34 \\
Low-rank weekly cycle & 1.23 & 3.91 & 11.85 \\
10-minute lag & \textbf{1.15} & \textbf{3.55} & \textbf{10.29} \\
-----------------------------\\
Static shortest path & 0.97 & 3.02 & 9.61 \\
\bottomrule
\end{tabular}
\end{table}

We further examined robustness across route complexity. Figure 2 shows that 90th percentile regret grows with path length for all predictors, but our cyclostationary models maintain a narrow gap with near-real-time benchmarks, while lagged baselines diverge substantially for longer routes.

\begin{figure}[!t]
    \centering
    \includegraphics[width=0.9\columnwidth]{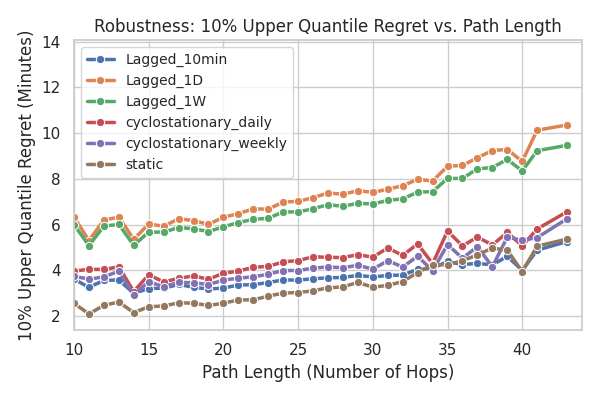}
    \caption{10\% upper quantile regret as a function of path length (number of hops in the unweighted shortest path per OD pair). Only path lengths with more than 1,000 samples are shown to ensure stable quantile estimates.}
    \label{fig:regret}
\end{figure}

\section{Conclusions }
We introduced a large-scale, year-long dataset of travel times over ~5,000 arterial road segments in Seoul. We proposed a low-rank cyclostationary forecasting model trained on this data. Remarkably, our results revealed that even predictors with access to near-real-time experience significant regret in rare cases, indicating that the remaining errors stem from inherently unpredictable traffic fluctuations. Within this practical ceiling, our model requires only data from the preceding week and still achieves a mean regret of 1.23 minutes on long routes. Moreover, the tail-regret closely matches a 10-minute-lag dynamic benchmark. These encouraging findings suggest that traffic planning can rely on accurate predictions computed a week before deployment. Given the model’s slow week-to-week variation, it is reasonable to expect that similar performance may extend to longer forecasting horizons, although this was not explicitly evaluated. 

Our simple predictor was sufficient to achieve near-real-time performance using a spatial basis estimated once from initial data. However, demand patterns may vary due to urban development, shifting the underlying spatial structure of the network over time. Future work will explore adaptive estimation of the low-rank spatial basis to track such structural drift as it occurs.
\section{Supplementary material}
We conducted more experiments to validate the sensitivity to assumptions.
We first provide a study of model order selection. Then we demonstrate the temporally cyclic structure of the spatial eigenmodes with multiple cycles (daily and weekly) and finally we provide a comparison to predictors based on cylclostationarity (full rank) and low rank only (no periodicity assumptions) 
\subsection*{A. MDL Model Order Selection}
\begin{figure}[h]
    \centering
    \includegraphics[width=0.8\columnwidth]{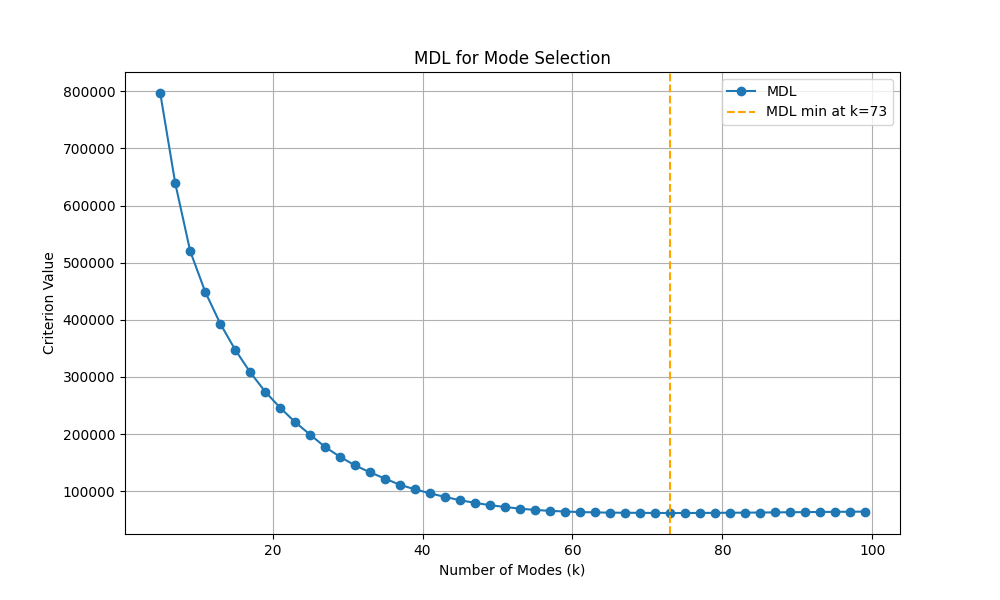}
    \caption{MDL criterion vs.\ number of modes $k$. The minimum is at $k=73$ (dashed line), but the curve nearly plateaus around $k=25$.}
    \label{fig:mdl}
\end{figure}

To test the model order, we applied the MDL principle \cite{wax1985detection} 
to the spatio-temporal traffic matrix. 
Figure~\ref{fig:mdl} shows the Minimum Description Length (MDL) criterion as a function of the number of retained modes $k$. The MDL minimum occurs at $k=73$. However, the criterion nearly flattens around $k=25$, indicating that the additional modes beyond 25 contribute noise components only. Hence, to reduce complexity, we chose $k=25$ in the main paper.

\subsection*{B. Power Spectral Density of the modes}
\begin{figure}[h]
    \centering
  \includegraphics[width=0.8\columnwidth]{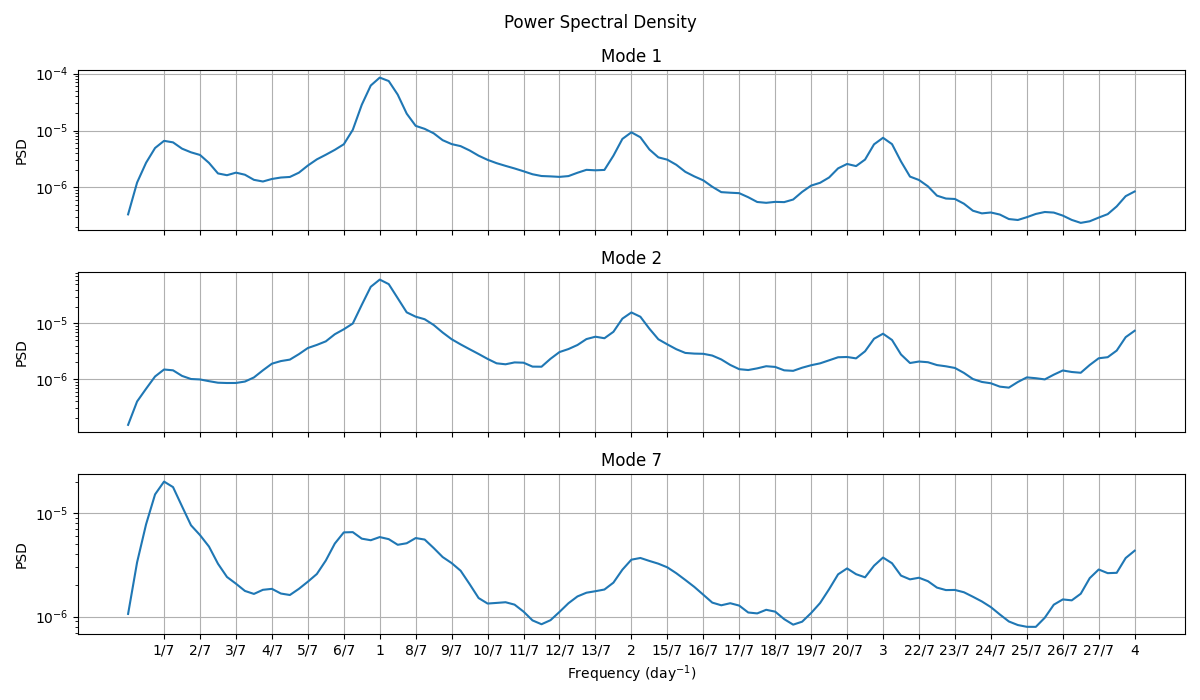}
    \caption{Power spectral density of temporal singular vectors. Modes 1 and 2 are dominated by daily periodicity ($1\,\text{day}^{-1}$), while Mode 7 exhibits a stronger weekly component ($1/7\,\text{day}^{-1}$).}
    \label{fig:psd}
\end{figure}

Figure~\ref{fig:psd} presents the estimated power spectral density (PSD) of selected right singular vectors from the truncated SVD, computed using the Welch periodogram 
\cite{stoica2005spectral} 
with Hann window, sampling frequency of 144 samples/day, and FFT size 144*28. Modes 1 and 2 exhibit a dominant peak at the daily frequency ($1\,\text{day}^{-1}$), with secondary peaks at the weekly frequency ($1/7\,\text{day}^{-1}$) and their harmonics. Mode 7 demonstrates the opposite pattern: the weekly frequency dominates while the daily peak is attenuated, indicating that certain spatial patterns are governed primarily by weekly periodicity. Together, these modes confirm the low-rank cyclostationary structure exploited by our predictor.
\subsection*{Comparison to full rank cyclic prediction and to low-rank non-periodic analysis}
We now compare the proposed predictor to the full-rank cyclic predictor and to the low-rank predictor without the cyclic structure. We also utilized the MDL model order selection to justify the selection of $k=25$.
The \emph{cyclostationary-only} variants use the cyclostationary temporal model with a full-rank ($k=4729$), while the \emph{low-rank static} variant applies the low-rank spatial basis without using the cyclostationary dynamics. 
The results are presented in Table~\ref{tab:ablation}. 
Model order $k=73$ yields the best resultsbut these are almost identical to $k=25$. 
The full-rank model performs slightly worse because of the effect of added noise from the higher modes. However, it has the best $1\%$ worst case, but the difference is non-substantial.  
\begin{table}[!t]
\centering
\caption{Ablation Study: Regret Statistics (Minutes) vs.\ Dynamic Real-Time}
\label{tab:ablation}
\begin{tabular}{lccc}
\toprule
\textbf{Predictor}
  & \textbf{Mean}
  & \shortstack{\textbf{10\% upper} \\ \textbf{quantile}}
  & \shortstack{\textbf{1\% upper} \\ \textbf{quantile}} \\
\midrule
Low-rank w.o. cyclostationarity & 2.17 &  6.39 & 19.44 \\
Low-rank (25) daily cycle   & {1.38} & 4.31 & 12.34 \\
Low-rank (25) weekly cycle  & {1.23} & 3.91 & 11.85 \\
Low-rank (73) weekly cycle  & \textbf{1.22} & {\bf 3.89} & {\bf 11.77} \\
Full-rank daily cycle  & 1.31 &  4.14 & 11.92 \\
Full-rank weekly cycle & 1.25 &  3.96 & 11.80 \\
\bottomrule
\end{tabular}
\end{table}

\clearpage


\begin{thebibliography}{10}
\providecommand{\url}[1]{#1}
\csname url@samestyle\endcsname
\providecommand{\newblock}{\relax}
\providecommand{\bibinfo}[2]{#2}
\providecommand{\BIBentrySTDinterwordspacing}{\spaceskip=0pt\relax}
\providecommand{\BIBentryALTinterwordstretchfactor}{4}
\providecommand{\BIBentryALTinterwordspacing}{\spaceskip=\fontdimen2\font plus
\BIBentryALTinterwordstretchfactor\fontdimen3\font minus \fontdimen4\font\relax}
\providecommand{\BIBforeignlanguage}[2]{{%
\expandafter\ifx\csname l@#1\endcsname\relax
\typeout{** WARNING: IEEEtran.bst: No hyphenation pattern has been}%
\typeout{** loaded for the language `#1'. Using the pattern for}%
\typeout{** the default language instead.}%
\else
\language=\csname l@#1\endcsname
\fi
#2}}
\providecommand{\BIBdecl}{\relax}
\BIBdecl

\bibitem{kim2005optimal}
S.~Kim, M.~E. Lewis, and C.~C. White, ``Optimal vehicle routing with real-time traffic information,'' \emph{IEEE Transactions on Intelligent Transportation Systems}, vol.~6, no.~2, pp. 178--188, 2005.

\bibitem{lee2024transfer}
D.~Lee, ``Transfer learning-based deep reinforcement learning approach for robust route guidance in mixed traffic environment,'' \emph{IEEE Access}, vol.~12, pp. 61\,667--61\,680, 2024.

\bibitem{lagos2025online}
T.~Lagos, R.~Auad, and F.~Lagos, ``The online shortest path problem: Learning travel times using a multiarmed bandit framework,'' \emph{Transportation Science}, vol.~59, no.~1, pp. 28--59, 2025.

\bibitem{chen2013finding}
B.~Y. Chen, W.~H. Lam, A.~Sumalee, Q.~Li, H.~Shao, and Z.~Fang, ``Finding reliable shortest paths in road networks under uncertainty,'' \emph{Networks and spatial economics}, vol.~13, no.~2, pp. 123--148, 2013.

\bibitem{zhang2023finding}
Z.~Zhang and M.~Li, ``Finding paths with least expected time in stochastic time-varying networks considering uncertainty of prediction information,'' \emph{IEEE Transactions on Intelligent Transportation Systems}, vol.~24, no.~12, pp. 14\,362--14\,377, 2023.

\bibitem{falek2022re}
A.~M. Falek, A.~Gallais, C.~Pelsser, S.~Julien, and F.~Theoleyre, ``To re-route, or not to re-route: Impact of real-time re-routing in urban road networks,'' \emph{Journal of Intelligent Transportation Systems}, vol.~26, no.~2, pp. 198--212, 2022.

\bibitem{asif2016matrix}
M.~T. Asif, N.~Mitrovic, J.~Dauwels, and P.~Jaillet, ``Matrix and tensor based methods for missing data estimation in large traffic networks,'' \emph{IEEE Transactions on intelligent transportation systems}, vol.~17, no.~7, pp. 1816--1825, 2016.

\bibitem{mitrovic2015low}
N.~Mitrovic, M.~T. Asif, J.~Dauwels, and P.~Jaillet, ``Low-dimensional models for compressed sensing and prediction of large-scale traffic data,'' \emph{IEEE Transactions on Intelligent Transportation Systems}, vol.~16, no.~5, pp. 2949--2954, 2015.

\bibitem{lin20243d}
M.~Lin, J.~Liu, H.~Chen, X.~Xu, X.~Luo, and Z.~Xu, ``A 3d convolution-incorporated dimension preserved decomposition model for traffic data prediction,'' \emph{IEEE Transactions on Intelligent Transportation Systems}, 2024.

\bibitem{wang2021graphtte}
Q.~Wang, C.~Xu, W.~Zhang, and J.~Li, ``Graphtte: Travel time estimation based on attention-spatiotemporal graphs,'' \emph{IEEE Signal Processing Letters}, vol.~28, pp. 239--243, 2021.

\bibitem{li2024sts}
L.~Li, K.~Yang, J.~Bi, and F.~Luo, ``{STS-CCL}: {S}patial-temporal synchronous contextual contrastive learning for urban traffic forecasting,'' in \emph{IEEE International Conference on Acoustics, Speech and Signal Processing (ICASSP)}.\hskip 1em plus 0.5em minus 0.4em\relax IEEE, 2024, pp. 6705--6709.

\bibitem{jin2024spatio}
J.~Jin, J.~Zhang, J.~Tang, S.~Liang, and Z.~Qu, ``Spatio-temporal data mining with information integrity protection: Graph signal based air quality prediction,'' in \emph{IEEE International Conference on Acoustics, Speech and Signal Processing (ICASSP)}.\hskip 1em plus 0.5em minus 0.4em\relax IEEE, 2024, pp. 5190--5194.

\bibitem{zhang2024arfa}
W.~Zhang, X.~Zou, L.~Wu, X.~Wang, J.~Huang, and J.~Xing, ``Arfa: An asymmetric receptive field autoencoder model for spatiotemporal prediction,'' in \emph{IEEE International Conference on Acoustics, Speech and Signal Processing (ICASSP)}.\hskip 1em plus 0.5em minus 0.4em\relax IEEE, 2024, pp. 3230--3234.

\bibitem{ermagun2018spatiotemporal}
A.~Ermagun and D.~Levinson, ``Spatiotemporal traffic forecasting: review and proposed directions,'' \emph{Transport Reviews}, vol.~38, no.~6, pp. 786--814, 2018.

\bibitem{yu2017spatio}
B.~Yu, H.~Yin, and Z.~Zhu, ``Spatio-temporal graph convolutional networks: A deep learning framework for traffic forecasting,'' \emph{arXiv preprint arXiv:1709.04875}, 2017.

\bibitem{xu2025fddsgcn}
J.~Xu, C.~Zhao, J.~Yang, Y.~Huang, Y.~Yang, and L.~Yee, ``Fddsgcn: Fractional decoupling dynamic spatiotemporal graph convolutional network for traffic forecasting,'' in \emph{IEEE International Conference on Acoustics, Speech and Signal Processing (ICASSP)}.\hskip 1em plus 0.5em minus 0.4em\relax IEEE, 2025, pp. 1--5.

\bibitem{wu2025sfadnet}
M.~Wu, W.~Weng, J.~Li, Y.~Lin, J.~Chen, and D.~Seng, ``Sfadnet: Spatio-temporal fused graph based on attention decoupling network for traffic prediction,'' in \emph{IEEE International Conference on Acoustics, Speech and Signal Processing (ICASSP)}.\hskip 1em plus 0.5em minus 0.4em\relax IEEE, 2025, pp. 1--5.

\bibitem{guo2019deep}
S.~Guo, Y.~Lin, S.~Li, Z.~Chen, and H.~Wan, ``Deep spatial--temporal {3D} convolutional neural networks for traffic data forecasting,'' \emph{IEEE Transactions on Intelligent Transportation Systems}, vol.~20, no.~10, pp. 3913--3926, 2019.

\bibitem{wax1985detection}
M.~Wax and T.~Kailath, ``Detection of signals by information theoretic criteria,'' \emph{IEEE Transactions on Acoustics, Speech, and Signal Processing}, vol.~33, no.~2, pp. 387--392, 1985.

\bibitem{stoica2005spectral}
P.~Stoica, R.~L. Moses \emph{et~al.}, \emph{Spectral analysis of signals}.\hskip 1em plus 0.5em minus 0.4em\relax Pearson Prentice Hall Upper Saddle River, NJ, 2005, vol. 452.

\bibitem{openstreetmap}
{OpenStreetMap contributors}, ``Openstreetmap,'' Available: \url{https://www.openstreetmap.org}, 2024, accessed: July. 15, 2025.

\bibitem{boeing2017osmnx}
G.~Boeing, ``Osmnx: New methods for acquiring, constructing, analyzing, and visualizing complex street networks,'' \emph{Computers, Environment and Urban Systems}, vol.~65, pp. 126--139, 2017.

\bibitem{blondel2008fast}
V.~D. Blondel, J.-L. Guillaume, R.~Lambiotte, and E.~Lefebvre, ``Fast unfolding of communities in large networks,'' \emph{Journal of Statistical Mechanics: Theory and Experiment}, vol. 2008, no.~10, p. P10008, 2008.
\end{thebibliography}
\end{document}